\documentclass[11pt,a4paper]{article}

\usepackage{amsmath,amssymb,amsfonts}
\usepackage[dvips]{graphicx}
\usepackage{hyperref}
\hypersetup{
    colorlinks=true,
    linkcolor=blue,
    citecolor=blue,
    urlcolor=blue,
    pdftitle={Title of the Paper},
    pdfauthor={Author Name}
}

\usepackage{geometry}
\usepackage[utf8]{inputenc}
\usepackage{color, float}
\usepackage{subfigure}
\usepackage{url}
\usepackage{authblk}
\usepackage{array}
\usepackage{slashed}
\usepackage[svgnames]{xcolor} 
\usepackage{tikz-feynman}
\tikzfeynmanset{/tikzfeynman/momentum/arrow distance=2mm,
/tikzfeynman/momentum/arrow shorten=0.25}
\usepackage[small,bf]{caption}  
\usepackage{microtype}
\usepackage[bottom,para]{footmisc}
\usepackage[backend=biber]{biblatex}
\usepackage{booktabs}         

\usepackage{authblk}

\renewcommand{\Im}{{\rm Im}}

\def\beq {\begin{equation}}
\def\eeq {\end{equation}}
\def\bea {\begin{eqnarray}}
\def\eea {\end{eqnarray}}

\title{\bf \boldmath Probing the structure of the $B$ meson with $B\to\ell\ell\ell'\nu$ }

\author{Aoife Bharucha$^1$\thanks{aoife.bharucha@cpt.univ-mrs.fr},  Bharti Kindra$^2$\thanks{bharti.kindra04@gmail.com}, Namit Mahajan$^3$\thanks{nmahajan@prl.res.in}\\
	$^1$Aix Marseille Univ, Universit\'{e} de Toulon, CNRS, CPT, Marseille, France\\
	$^2$Indian Institute of Technology, Gandhinagar, India\\
	$^3$ Physical Research Laboratory, Ahmedabad, India    }  
\date{\today}

\begin{document}
\begingroup
\let\newpage\relax
\begin{flushright}INT-PUB-21-003\end{flushright}
\maketitle
\endgroup

\begin{abstract}
We consider the decay $B\to\ell\ell\ell^{\prime}\nu$ in QCD factorization for the case of non-zero (but small) $q^2$, taking into account leading power contributions at NLL in $\mathcal{O}(\alpha_s)$ and next-to-leading power corrections at $\mathcal{O}(\alpha_s^0)$. We further incorporate soft contributions obtained from light-cone sum rules, extrapolated from the Euclidean to physical region using once-subtracted dispersion relations. We restrict ourselves to the case $\ell\neq\ell'$ as otherwise the same sign $\ell$ and $\ell'$ cannot be distinguished.  We study the sensitivity of the results to the leading moment of the $B$-meson distribution amplitude and discuss the potential to extract this quantity at LHCb and the Belle II experiment.
\end{abstract}

	\section{Introduction}
\label{sec:intro}	
	Our understanding of the structure of the $B$ meson lacks the precision needed to provide accurate predictions of $B$-meson decays beyond naive factorization.
The key non-perturbative input entering both QCD factorization and light-cone sum rules (LCSR) is the leading inverse moment of the $B$-meson distribution amplitude, $\lambda_B$, defined by (see e.g.~Ref.~\cite{Braun:2003wx})
	\begin{equation}
	\label{eq:lambdaB}
	\lambda_B^{-1}(\mu)=\int_{0}^{\infty} \frac{dk}{k}\phi_B^{+}(k,\mu),
	\end{equation} 
	where $\phi_B^+(k,\mu)$ is the distribution amplitude of the $B$ meson, depending on the scale $\mu$.
The theoretical uncertainty on $\lambda_B$  is large with estimates ranging from $200$ MeV obtained using non-leptonic decays \cite{Beneke:2009ek,Beneke:2003zv} to $460\pm 110$ MeV obtained from QCD sum rules \cite{Braun:2003wx}. 

It was pointed out in Refs.~\cite{Beneke:2000wa,Grozin:1996pq,Bosch:2003fc} to use the charged current decay of the $B^+$ meson, with an extra photon in the final state i.e., $B^+ \to \ell^+\nu\gamma$ to probe $\lambda_B$ in the kinematic limit when the energy of the photon is large in comparison to the scale of strong interactions ($\Lambda_{\rm QCD}$). Note that the branching ratio is larger than for the purely leptonic final state, as the emission of the photon lifts the helicity suppression. On the experimental side, the most stringent limit from the BaBar collaboration was~\cite{Aubert:2009ya}
	\begin{equation}
	\mathcal{B}(B^+\to \ell^+\nu\gamma)< 15.6 \times 10^{-6},
	\end{equation}
	at $90\%$ confidence level (C.L.), resulting in the lower limit $\lambda_B>300$ MeV \cite{Aubert:2009ya}.
The Belle collaboration has provided an upper limit on the branching ratios for the electron and muon final states~\cite{Heller:2015vvm}:
	\begin{align}
	\mathcal{B}(B^+\to e^+\nu\gamma)&< 4.3 \times 10^{-6},\\
	\mathcal{B}(B^+\to \mu^+\nu\gamma)&< 3.4 \times 10^{-6},
	\end{align}
	also at $90\%$ C.L.,
	providing a lower limit of $\lambda_B>238$ MeV. Note that the fact that the Belle lower limit lies below the BaBar limit, despite the limit on the branching ratio being more stringent, is due both to a different choice of input parameters (particularly $m_b$ and $|V_{ub}|$) and the fact that in Ref.~\cite{Heller:2015vvm} the expression for the partial branching fraction includes state-of-the-art next-to-leading order corrections from Ref.~\cite{Beneke:2011nf} and the soft corrections calculated in LCSR from Ref.~\cite{Braun:2012kp}.

Updating this result is a priority at Belle II, and projections look very promising~\cite{Gelb:2018end,Kou:2018nap}. However, the measurement of $B\to\gamma\ell\nu$  at LHCb is challenging as the low energy photon in combination with the neutrino in the final state provide complications for the trigger. On the other hand, if the photon was off-shell and emitted a lepton pair, then for the final state consisting of three charged leptons and a neutrino  the analysis would be feasible. In fact, an upper limit at 95\% C.L. has already been provided by LHCb in the region where the lowest of the muon pair mass combination is below $980$ MeV \cite{Aaij:2018pka}, 
	\begin{equation}
	\mathcal{B}(B^+\to \mu^+ \mu^- \mu^+ \nu)<1.6 \times 10^{-8}.
	\end{equation}
In light of this measurement, it is of interest to assess whether
$B\to\ell\ell\ell'\nu$ could provide complementary information on the $B$-meson distribution amplitude. On the theoretical side, significant progress has been made since the first version of this work appeared. The complete QCD factorization framework for $B\to\ell\ell\ell'\nu$ at low $q^2$ was developed in Ref.~\cite{Beneke:2021rjf}, including leading power corrections at $\mathcal{O}(\alpha_s)$ and next-to-leading power corrections at  $\mathcal{O}(\alpha_s^0)$. Here a Breit Wigner ansatz was used for the soft corrections. This framework was subsequently extended through the derivation of additional subleading-power factorization formulae and a detailed phenomenological analysis of angular observables in Ref.~\cite{Wang:2021yrr}. Complementary theoretical descriptions have also been developed. The case of identical leptons in the final state was analysed in Ref.~\cite{Ivanov:2022uum}, a dispersive treatment of the form factors was presented in Ref.~\cite{Ivanov:2021jsr}, while LCSR predictions for the form factors were obtained in Ref.~\cite{Janowski:2021yvz}.

More recently, Ref.~\cite{Kurten:2022zuy} introduced a form-factor basis free from kinematic singularities. Within this framework, the soft contributions to the $B\to\gamma^*$ form factors were revisited in Ref.~\cite{Bharucha:2026bwx}. There, the QCD factorization results were extended to non-zero $q^2$, including the leading power corrections in $1/m_b$ and $1/v\cdot q$ and shown to admit a dispersive representation. This allowed the soft corrections to be calculated at $q^2\leq 0$ using the LCSR set-up of Ref.~\cite{Braun:2012kp}, and were shown to be less than $\sim 10\%$. The present version has therefore been updated accordingly. In particular, we include the previously omitted longitudinal form factor, implement the correct renormalization-scale dependence following Ref.~\cite{Beneke:2021rjf}, and replace
the previous treatment of the soft contributions by a once-subtracted dispersion-relation approach. Note that this is the first phenomenological analysis of the form factors and branching ratios, where the form factors are calculated in QCD factorization with leading power corrections at NLL and next-to leading power corrections at  $\mathcal{O}(\alpha_s^0)$, adding the
soft corrections calculated using a once-subtracted dispersion relation where the subtraction point is given in the Euclidean by the LCSR set-up of Ref.~\cite{Braun:2012kp}. The resulting framework provides a consistent description of the low-$q^2$ region while leaving the phenomenological conclusions of the original analysis qualitatively unchanged.

Our aim is therefore to provide predictions for the differential decay spectrum and the partial branching fraction for $B\to\ell\ell\ell^{\prime}\nu$, which, with more data, could lead to the measurement of the first moment of the $B$-meson distribution amplitude at LHCb. This would serve as an important cross-check for the results from Belle II. Our calculation is performed within the QCD factorization and LCSR framework and is therefore restricted to low $q^2$, where $q^2$ is the di-lepton mass squared. Since when $\ell=\ell^{\prime}$, $q^2$ cannot be uniquely defined experimentally such that the existing limit from LHCb cannot be converted into a lower limit on $\lambda_B$. We will therefore concentrate on the case $\ell\neq\ell^{\prime}$.

In the following section we will introduce our theoretical framework, expressing the amplitude for the decay in terms of the form factors, and providing details of the various contributions to these form factors that we include. In Sec.~\ref{sec:results}, after having discussed the kinematics necessary to calculate the branching ratio and the parameters adopted, we will present the results of our numerical analysis. Finally, a discussion of these results and our conclusions can be found in Sec.~\ref{sec:conclusion}.

\section{Theoretical framework}
	\subsection{The Effective Amplitude}\label{4lep_amplitude}
The amplitude for the process $B^+(p_B)\to \ell^+(q_1)\ell^-(q_2) \ell^{\prime +}(p_1) {\nu}(p_2)$ can be written as
\begin{equation}\label{eq:ampli}
iA=-\frac{G_F V_{ub}}{\sqrt{2}}\frac{ie^2}{q^2}(\bar{u}_{\ell}\gamma^\mu v_\ell)[(\bar{u}_\nu \gamma^\rho P_Lv_{\ell^{\prime}})T_{\mu\rho}-i(\bar{u}_\nu \gamma_\mu P_Lv_{\ell^{\prime}})f_B]
\end{equation}
 where $G_F$ is the Fermi constant, $V_{ub}$ is the CKM matrix element, $e$ is the electric charge and $P_L=1-\gamma_5$. Here the first term describes the emission of the virtual photon from the $B$ meson and the second term from the lepton.  At leading order the second term can trivially be written in terms of the $B$-meson decay constant $f_B$, whereas the first term is more complicated and requires further study. On writing down this term in the most general form as possible, imposing the conservation of the electromagnetic current, applying the equation of motion (see Appendix \ref{AppendixA}) and neglecting the lepton masses, the tensor $T_{\mu\rho}$ defined in Eq.~\eqref{eq:ampli} reduces to,
 \begin{equation}
T_{\mu\rho}=i F_A(g_{\mu\rho}\,v\cdot q-v_\mu q_\rho)+F_V\epsilon_{\rho\mu\lambda\sigma} v^\lambda q^\sigma-i F_{\parallel}v_\mu q_\rho,
\label{eq:Tmunu}
\end{equation}
where $p_B=m_B v$ and $q=q_1+q_2$ (in the following we will also require $p=p_1+p_2$).
Making predictions for these decays therefore comes down to obtaining expressions for these form factors.
Note that as the mass of the leptons in the final state has been neglected, there is an exact cancellation of the contact term with the contribution of photon emission from charged lepton. For the case of tau leptons in the final state, additional terms may appear and would need to be calculated.

	\subsection{Form Factors}
	
The form factors can be factorized in the limit that the quark propagator between the weak and electromagnetic vertices is far off-shell. This occurs for example when the quark propagator is hard-collinear, which is case for $q^2\ll m_B^2$, where the leading-power result can be obtained by matching QCD to SCET I, and then to HQET. 
The form factors  have been calculated in this way for the case $B\to\gamma\ell\nu$ (i.e.~$q^2=0$). We will therefore first review the results for $q^2=0$, before presenting our calculation at non-zero $q^2$.

\subsubsection{State-of-the-art result for $B\to\gamma\ell\nu$}
 In the case of $B\to\gamma\ell\nu$, from Ref.~\cite{Beneke:2018wjp} we have
 \begin{align}
\nonumber  F_V=&\frac{Q_u\,f_B\,m_B}{2\,E_\gamma\,\lambda_B(\mu)}R(E_\gamma\,\mu)+\xi(E_\gamma)+\Delta\xi(E_\gamma)\\
\label{eq:FVFA}  F_A=&\frac{Q_u\,f_B\,m_B}{2\,E_\gamma\,\lambda_B(\mu)}R(E_\gamma,\mu)+\xi(E_\gamma)-\Delta\xi(E_\gamma),
 \end{align}
 where the photon energy $E_\gamma=(m_B^2-p^2)/(2 m_B)$, $Q_u$ is the charge of the $u$ quark and $m_B$ is the mass of the $B$ meson.
 The first term is the leading contribution in the heavy quark expansion, which depends inversely on the quantity of interest, $\lambda_B$, the first inverse moment of the $B$-meson distribution amplitude defined in Eq.~\eqref{eq:lambdaB}. 
Note here that the factor $R(E_\gamma,\mu)$ contains the radiative corrections calculated in Ref.~\cite{Beneke:2011nf}, and at tree level is equal to 1. 
The second term encodes the symmetry conserving soft and $\mathcal{O}(1/m_b,1/E_\gamma)$ corrections to the form factors which may be sizeable. The soft corrections have been calculated in LCSR up to next-to-leading order in $\alpha_s$ (NLO) at leading twist and up to twist-6 at leading order in $\alpha_s$~\cite{Beneke:2018wjp,Wang:2016qii,Wang:2018wfj}.
The last term contains the symmetry breaking corrections, which arise at $\mathcal{O}(1/m_b,1/E_\gamma)$ in QCD factorization, as well as receiving contributions from the  soft corrections at twist-3 to 6~\cite{Wang:2016qii,Wang:2018wfj,Beneke:2018wjp}. 
	
	\begin{figure}[tb]
		\begin{center}
			\begin{tabular}{cc}
 \begin{tikzpicture}
\begin{feynman}[large]
\vertex (a);
\vertex [below=of a] (b);
\vertex [left=of a] (i1) {\(b\)};
\vertex [right=of a] (f1){\(W^-\)};
\vertex [left=of b] (i2) {\(\bar{u}\)};
\vertex [right=of b] (f2) {\(\gamma^*\)};
\diagram* {
(i1) -- [ momentum=\(p_B-k\)]  (a) -- (b),(i2) -- [ momentum'=\(k\)]  (b),
(a) -- [boson, momentum=\(p\)] (f1),
(b) -- [boson, momentum'=\(q\)] (f2)
};
\end{feynman}
\end{tikzpicture} &  \begin{tikzpicture}
\begin{feynman}[large]
\vertex (a);
\vertex [below=of a] (b);
\vertex [left=of a] (i1) {\(b\)};
\vertex [right=of a] (f1){\(\gamma^*\)};
\vertex [left=of b] (i2) {\(\bar{u}\)};
\vertex [right=of b] (f2) {\(W^-\)};
\diagram* {
(i1) -- [ momentum=\(p_B-k\)]  (a) -- (b),(i2) -- [ momentum'=\(k\)]  (b),
(a) -- [boson, momentum=\(q\)] (f1),
(b) -- [boson, momentum'=\(p\)] (f2)
};
\end{feynman}
\end{tikzpicture}
\\[.3cm]
 (a) & (b)
			\end{tabular}	\caption{\label{tree} Feynman diagrams showing (a) the emission of a photon from the $u$ quark, (b) photon emission from the $b$ quark.}
		\end{center}
	\end{figure}
	
	\subsubsection{Symmetric corrections for $B\to \ell\ell\ell'\nu$ at leading order in $\alpha_s$}
Having studied the form factors for on-shell photons, we now wish to extend these to the case of non-zero $q^2$. In $B\to\gamma\ell\nu$, factorization holds
as long as the photon is energetic in the rest frame of the $B$ meson, implying that the photon energy $\sim m_b$. 
Similarly in our case a hard collinear virtual photon coupling to the spectator $u$ quark can be integrated out systematically, and the total amplitude is found to factorize into a hard scattering kernel and a soft part, given by the $B$ meson distribution 
amplitude. In the same setting, the resulting contribution from the $b$ quark is found to be power suppressed. A quick analysis
of the diagrams for the case of a soft emitted photon reveals that there are possibly power enhanced contributions, coming
both from the $u$-quark and $b$-quark legs. However, these are not expected to be factorizable and will not be
considered any further.
We thus restrict our attention to the case of a hard collinear photon (defined below) for which the factorization is found to
hold, constraining $q^2$ to be not too far from zero.
Thus at the leading order, only the diagram in Fig.~\ref{tree} (a) contributes to the form factors.

In order to study this diagram in more detail, for convenience, we choose to work in light-cone coordinates ($l\equiv (l_+,l_-,l_{\perp})$) where, 
	\begin{align}
	l_{\pm}=\frac{l_0\pm l_3}{2},\,\quad l_{\perp}=(l_1,l_2),\,\quad
	\mbox{and}\qquad l^\mu=\frac{l_+}{2} n_-^\mu+l_\perp+\frac{l_-}{2} n_+^\mu.
	\end{align}
For the required case of a hard collinear virtual photon, the momentum $q$ will scale as $q=(q_+,q_-,q_{\perp})\sim(\lambda,1,\lambda^{1/2})$, and the momentum of the soft spectator quark will scale as $k=(k_+,k_-,k_{\perp})\sim (\lambda,\lambda,\lambda)$. This implies that,
	\begin{equation}
(q-k)^2\sim q^2-q_-k_+\ldots
	\end{equation}
where we neglect terms that are suppressed by higher powers of $\lambda$, such that the propagator can be expressed as
\begin{equation}\label{eq:prop}
\frac{\slashed{q}-\slashed{k}}{(q-k)^2}=	\frac{q_-\slashed{n}_+/2}{q^2-q_-k_+}-\left(	\frac{k_+\slashed{n}_-/2}{q^2-q_-k_+}+	\frac{k_-\slashed{n}_+/2}{q^2-q_-k_+}+	\frac{\slashed{k}_\perp}{q^2-q_-k_+}\right).
\end{equation}
Here the first term provides the leading contribution and the terms in brackets are suppressed by $\lambda$.
Using this expansion of the propagator, and following Ref.~\cite{Braun:2012kp}, the form factor at leading order in the heavy quark and the perturbative expansion, arising from the first term in Eq.~\eqref{eq:prop}, reads,
	\begin{equation}\label{eq:FVFALO}
F^{\rm LO}_{V/A}=Q_u f_B\, m_B\int_{0}^{\infty} dk_+ \frac{\phi_B^+(k_+)}{q_-\,k_+-q^2-i\epsilon}.
	\end{equation}
We now wish to include the soft contribution, which was calculated for the case of $B\to\gamma\ell\nu$ using dispersion relations and quark-hadron duality in Refs.~\cite{Braun:2003wx,Beneke:2018wjp} using a technique similar to that applied to the $\gamma^*\gamma\pi$ form factor~\cite{Khodjamirian:1997tk}. The idea  is to relate via a dispersion relation the form factor at $q^2=0$ to that at asymptotically large values of $-q^2$, where the result can be calculated perturbatively to arbitrary precision in the expansion in $1/m_b$, $q^2$. This requires an assumption to be made about the hadronic spectral density as a function of $q^2$ (for $q^2$ in the Euclidean). 
Following Refs.~\cite{Braun:2012kp,Beneke:2018wjp}, one easily obtains the result for the two form factors $F_V$ and $F_A$ at leading order in the twist and perturbative expansion (for $q^2\leq 0$),
	\begin{align}\label{eq:formfac}
F^{\rm LO}_{V/A}+\xi^{\rm Eucl.}_{\rm soft}(q^2)= &\,Q_u f_B\,m_B\bigg( \int_{s_0/q_-}^{\infty}dk_+ \frac{\phi_B^+(k_+)}{q_-\,k_+-q^2-i\epsilon}\nonumber\\
	 &\qquad \qquad\qquad+ \int_0^{s_0/q_-} dk_+ \frac{e^{-(q_-k_+-m_{\rho}^2)/M^2}}{m_{\rho}^2-q^2-i \epsilon} \phi_B^+(k_+)\bigg).
	\end{align}
	where $s_0$ is the continuum threshold, $m_{\rho }$ is the mass of $\rho$ meson, and $M^2$ is the Borel parameter. Note that $s_0$ corresponds to the value of $q^2$ above which quark-hadron duality is expected to hold. The Borel parameter enters the result due to fact that a Borel transformation has been performed in order to reduce the sensitivity to this assumption. 
	By comparing Eqs.~\eqref{eq:FVFALO} and \eqref{eq:formfac}  the soft contribution at leading order to the form factor  in the Euclidean can be extracted,
\begin{equation}\label{eq:softLO}\hspace{-.2cm}
\xi^{\rm Eucl.}_{\rm soft}(q^2)=Q_u f_B\,m_B\int_{0}^{s_0/q_-}dk_+ \left( \frac{e^{-(q_-k_+-m_{\rho}^2)/M^2}}{m_{\rho}^2-q^2-i \epsilon}-\frac{1}{q_-\, k_+-q^2-i\epsilon}\right)\phi_B^+(k_+).
\end{equation}	 
However, we are in need of this soft contribution at $q^2>0$. This can be obtained by means of a once subtracted dispersion relation, as discussed in the following subsection.

\subsubsection{Symmetric corrections for $B\to \ell\ell\ell'\nu$ at NLL}
\label{sec:SymNLL}
Before coming to the once-subtracted dispersion relations, let us first consider the NLL corrections which were included for the on-shell photon case in Eq.~\eqref{eq:FVFA} via the factor $R(E_\gamma,\mu)$. The radiative corrections for $B\to \ell\nu\gamma$ at NLL are significant and were seen in Refs.~\cite{Beneke:2011nf,Beneke:2018wjp} to affect the form factors by $20-40\% $.  Note that these NLL corrections not only affect the leading-order contribution, i.e.~that given by Eq.~\eqref{eq:FVFALO}, but also the soft contribution, as described in Ref.~\cite{Beneke:2018wjp}. 
The factor $R(E_\gamma,\mu)$ introduced in Ref.~\cite{Beneke:2011nf} can be adapted to our case, 
 \begin{equation}\label{eq:rad}
R(q_-,q^2,\mu,\mu_{h1},\mu_{h2})= C(q_-,\mu_{h1})K^{-1}(\mu_{h2})U(q_-,\mu_{h1},\mu_{h2},\mu) J(q_-,q^2,\mu).
\end{equation}
Here $C(q_-,\mu_{h1})$ is obtained from matching the QCD heavy-to-light current to the corresponding SCET I current at the hard scale $\mu_{h1}$, the NLO result was calculated in Refs.~\cite{Lunghi:2002ju,Bosch:2003fc,Bauer:2000yr}. The expression for $C(E_\gamma,\mu_{h1})$ found in Ref.~\cite{Beneke:2011nf} for the $B\to\gamma\ell\nu$ case can be applied to $B\to\ell\ell\ell'\nu$ by replacing $2\,E_\gamma$ by $q_-$,
\begin{align}\hspace{-.2cm}
    C(q_-,\mu)=1-\frac{\alpha_s C_F}{4\pi}\bigg(&2 \ln^2 \frac{x}{\hat\mu}-5\ln \frac{x}{\hat\mu}+\frac{3-2 x}{1-x}\ln x+2\,\mathrm{Li}_2(1-x)+6
    +\frac{\pi^2}{12}\bigg),
\end{align}
where $x=q_-/m_b$ and $\hat\mu=\mu/m_b$.
The factor $K(\mu_{h2})$, which accounts for the conversion from the static $B$-meson decay constant in the SCET current to the standard definition in QCD, $f_B$, used in our analysis, can be directly applied to our case more details and the expression can be found in Ref.~\cite{Beneke:2011nf}. 

Letting aside momentarily the renormalization group evolution (RGE) factor $U(q_-,\mu_{h1},\mu_{h2},\mu)$, let us first discuss the factor $J(q_-,q^2,\mu)$. This accounts for the hard-collinear radiative corrections, and was calculated for $B\to\gamma\ell\nu$ in Refs.~\cite{Lunghi:2002ju,Bosch:2003fc}. For $B\to\ell\ell\ell'\nu$ where $q^2\neq 0$ we can adopt the expression given in Ref.~\cite{Wang:2016qii}.
\begin{align}\label{eq:Jfac}
\nonumber J(q_-,q^2,\mu)= 1+\frac{\alpha_s C_F}{4\pi}\bigg[&\ln^2\frac{\mu^2}{q_-\,\omega-q^2}-1-\frac{\pi^2}{6}\\
&-\frac{q^2}{q_-\,\omega}\ln\frac{q^2-q_-\,\omega}{q^2}\left(\ln\frac{\mu^2}{-q^2}+\ln\frac{\mu^2}{q_-\,\omega-q^2}+3\right)\bigg].
\end{align}
Returning to the RGE factor $U(q_-,\mu_{h1},\mu_{h2},\mu)$, we see that  for any choice of the scales $\mu_{h1}$, $\mu_{h2}$ and $\mu$, some large logarithms arising in the expressions for $C(q_-,\mu_{h1})$, $K(\mu_{h2})$ and $J(q_-,q^2,\mu)$ will remain. Therefore it makes sense to resum these logarithms to all orders by solving a renormalization group equation~\cite{Bosch:2003fc}. This resummation results in the factor $U(q_-,\mu_{h1},\mu_{h2},\mu)$ found in Ref.~\cite{Beneke:2011nf}, which again can be applied to $B\to\ell\ell\ell'\nu$ by replacing $2\,E_\gamma$ by $q_-$.
Note the dependence on three scales: the hard scales $\mu_{h1}$ and $\mu_{h2}$ are taken to be $q_-$ and $m_b$ respectively, while the hard-collinear scale $\mu$ is set to $(m_b\Lambda_{\rm QCD})^{1/2}$.

Having defined $R(q_-,q^2,\mu,\mu_{h1},\mu_{h2})$, we now have the result at leading order in the heavy quark expansion, and NLL in the perturbative expansion:
\begin{align}\label{eq:FVFA-NLL}
F_{V/A}^{\rm NLL}=&\,Q_u f_B\, m_B\int_0^\infty \frac{d\omega\,\phi_B^+(\omega,\mu)}{q_-\,\omega-q^2-i\varepsilon} R(q_-,q^2,\mu,\mu_{h1},\mu_{h2}),
\end{align}
\noindent where clearly these are symmetric contributions.
As mentioned earlier, including these NLL corrections has the consequence that the expression for soft contribution to the form factors for $q^2\leq 0$ given in Eq.~\eqref{eq:formfac} is no longer valid. We can derive the new expression for the soft corrections using the following relation~\cite{Beneke:2011nf},
\begin{equation}\label{eq:softNLL1}
\xi_{\rm soft}^{\rm Eucl.}(q^2)=\frac{1}{\pi}\int_0^{s_0}\mathrm{d}s \left(\frac{e^{-(s-m_\rho^2)/M^2}}{m_\rho^2-q^2-i\varepsilon} -\frac{1}{s-q^2-i\varepsilon}\right)\Im F^{\rm QCD}(E,s).
\end{equation}
Adopting $F^{\rm QCD}(E,s)=F^{\rm NLL}_{V/A}$ from Eq.~\eqref{eq:FVFA-NLL}, setting $\omega'=s/q_-$, and with the help of Appendix A of Ref.~\cite{Wang:2016qii}, we therefore obtain at NLL
\begin{align}
\nonumber\xi_{\rm soft}^{\rm Eucl.}(q^2)=\,&\,Q_u f_B\, m_B\, C(q_-,\mu_{h1})K^{-1}(\mu_{h2})U(q_-,\mu_{h1},\mu_{h2},\mu)\\
\label{eq:softNLL}& \int_0^{s_0/q_-}\mathrm{d}\omega' \left(\frac{e^{-(q_-\,\omega'-m_\rho^2)/M^2}}{m_\rho^2-q^2-i\varepsilon} -\frac{1}{q_-\,\omega'-q^2-i\varepsilon}\right)\phi_B^{+
\,\rm eff}(\omega',\mu)
\end{align}
 where $\phi_{B,{\rm eff}}^+(\omega,\mu)=\phi_{B}^+(\omega,\mu)+\delta\phi_{B,{\rm eff}}^+(\omega ',\mu)$, for $\delta\phi_{B,{\rm eff}}^+(\omega ',\mu)$ given by~\cite{Beneke:2018wjp,Wang:2016qii} \footnote{ The expression differs with Ref.~\cite{Wang:2016qii} as certain typographical errors have been corrected, after these corrections Eq.~\eqref{eq:deltaphi} agrees numerically exactly with Ref.~\cite{Beneke:2018wjp}.}
 \begin{align}
\nonumber \delta\phi_{B,{\rm eff}}^+(\omega',\mu)=\frac{\alpha_s(\mu) C_F}{4\pi}\bigg[&\int_0^{\omega'}d\omega\left(\frac{2}{\omega-\omega'}\ln\frac{\mu^2}{q_-(\omega'-\omega)}\right)_\oplus \phi_{B}^+(\omega,\mu)\\
\nonumber &-\omega'\int_0^{\omega'}d\omega\left(\frac{1}{\omega-\omega'}\ln\frac{\omega'-\omega}{\omega'}\right)_\oplus\frac{\phi_{B}^+(\omega,\mu)}{\omega}\\
\nonumber &+\frac{\omega'}{2}\int_0^\infty d\omega\ln^2\bigg|\frac{\omega-\omega'}{\omega'}\bigg|\frac{d}{d\omega}\frac{\phi_{B}^+(\omega,\mu)}{\omega}\\
\nonumber &-\int_{\omega'}^\infty d\omega\left(\ln^2\frac{\mu^2}{q_- \omega'}-\frac{\pi^2}{2}-1\right)\frac{d}{d\omega}\phi_{B}^+(\omega,\mu)\\
\nonumber&+\omega'\int_{\omega'}^\infty d\omega \bigg(\ln\frac{\mu^2}{q_-\omega'}\ln\frac{\omega-\omega'}{\omega'}-\frac{1}{2}\ln^2\frac{\mu^2}{q_-(\omega-\omega')}\\&+\frac{1}{2}\ln^2\frac{\mu^2}{q_-\omega'}
+3\ln\frac{\omega-\omega'}{\omega'}-\frac{2\pi^2}{3}\bigg)\frac{d}{d\omega}\frac{\phi_{B}^+(\omega,\mu)}{\omega}\bigg],\label{eq:deltaphi}
 \end{align}
 where the symbol $\oplus$ is defined by
 \begin{equation}
     \int_0^\infty d\omega\,\left(f(\omega,\omega')\right)_\oplus\, g(\omega)\equiv \int_0^\infty d\omega\, f(\omega,\omega')\left(g(\omega)-g(\omega')\right).
 \end{equation}
 The NLL expression for $\xi_{\rm soft}^{\rm Eucl.}$  is equivalent to Eq.~\eqref{eq:softLO}, and again we remind the reader that it is only valid at $q^2\leq 0$.
 
$\xi_{\rm soft}^{\rm Eucl.}$ can be extended to positive values of $q^2$ by means of a once subtracted dispersion relation, where close to the resonances i.e.~the $\rho$ and $\omega$ mesons, the non-perturbative contribution to the form factors can be obtained via vector meson dominance (VMD) arguments~\cite{Guadagnoli:2017quo,Kozachuk:2017mdk}: the
transition proceeds via $B\to V$, where $V$ is a vector meson which then converts into the virtual photon $\gamma^*$. The hadronic
correlator $T_{\mu\nu}$ is written by inserting a complete set of $1^{--}$ states, denoted by $\vert\rho\rangle$:
\begin{equation}
 T_{\mu\nu} = \int d^x e^{iqx}\langle \gamma^*(q)\vert J_{\mu}^{em}(x)\vert\rho(q)\rangle\langle\rho(q)\vert\bar{u}\gamma_{\nu}(1-\gamma_5)\vert B(p_B)\rangle,
\end{equation}
where $q$, the momentum of the virtual photon, is equal to that of the $\rho$ meson within VMD. VMD allows one to relate the photon and $\rho$ polarization vectors: $\epsilon_{\alpha}(\rho) = (e/g_{\gamma\rho})\epsilon_{\alpha}(\gamma)$,
with $e$ the electric charge and $g_{\gamma\rho}$ being the VMD coupling. Using the definition of the $\rho$ decay constant, $\langle 0\vert J_{\alpha}^{em}\vert\rho(q)\rangle = f_{\rho}m_{\rho}\epsilon_{\alpha}(\rho)$, and employing the standard definitions
of $B\to\rho$ form factors, $T_{\mu\nu}$ can be expressed in a way such that it can easily be compared with Eq.~\eqref{eq:Tmunu} in terms of $F_{A}$, $F_V$ and $F_{\parallel}$:
\begin{equation}
 T_{\mu\nu} = \frac{1}{2}\left(\frac{-if_{\rho}m_{\rho}}{q^2-m_{\rho}^2+im_{\rho}\Gamma_{\rho}}\right) X_{\mu\nu}
\end{equation}
where
\begin{align}
 X_{\mu\nu} = (m_B+m_{\rho})A_1(p^2)g_{\mu\nu}  +
 \frac{2V(p^2)}{m_B+m_{\rho}}\epsilon_{\nu\mu\rho\sigma}p_B^{\rho}q^{\sigma} - i \frac{2A_2(p^2)}{m_B+m_{\rho}}{p_B}_{\mu}q_{\nu}.
\end{align}
 The different form factors can then be readily identified. Particular attention needs to be paid to the term which 
is to be identified with $F_{\parallel}(p^2,q^2)$. When the photon is on shell, one requires $F_{\parallel}(p^2,q^2=0) = 0$ such that
the amplitude is described in terms of only two form factors, $F_A$ and $F_V$ as dictated by gauge invariance. Requiring
$F_{\parallel}(p^2,q^2=0) = 0$ and plugging the resulting relation back leads to the desired form of $F_{\parallel}(p^2,q^2)$:
\begin{equation}
 F_{\parallel} = -\frac{q^2}{v.q}\,\frac{m_B+m_\rho}{m_B^2-p^2} A_1(p^2),
\end{equation}
clearly vanishing for $q^2=0$.

Generalising the formalism to cover not only the case of the $\rho$ meson but a vector meson $V=\rho,\omega$, the once subtracted dispersion relation takes the form
\begin{equation}\label{eq:softdisp}
\xi_{X}^{\rm NLL}(q^2)=\xi^{\rm Eucl.}_{X,\,\rm soft}(q^2_0)+\sum_V\frac{q^2-q^2_0}{m_V^2-q^2_0}\frac{c_V f_V m_V  e^{i \delta_V}}{m_V^2-q^2-i m_V \Gamma_V}F^{B\to V}_X
\end{equation}
where $X=V,A$ and $\parallel$, $c_\rho=\tfrac12,\,\,c_\omega=\tfrac16$  and 
\begin{align}
\nonumber F^{B\to V}_V=&\frac{2m_B}{m_B+m_V} V^{B\to V}(p^2)\\
\nonumber F^{B\to V}_A=&\frac{m_B+m_V}{v.q} A_1^{B\to V}(p^2)\\
F^{B\to V}_\parallel=&-\frac{q^2}{v.q}\,\frac{m_B+m_V}{m_B^2-p^2} A_1^{B\to V}(p^2),
\end{align}
and $V^{B\to V}(p^2)$ and $A_1^{B\to V}(p^2)$ are the $B\to V$ transition form factors for $V=\rho,\omega$.
For $\xi_{V}^{\rm NLL}(q^2)$ and $\xi_{A}^{\rm NLL}(q^2)$, the subtraction constant is chosen to be the soft form factor in the Euclidean, where it is calculable, i.e.~$\xi^{\rm Eucl.}_{V/A,\,\rm soft}(q^2_0)=\xi^{\rm Eucl.}_{\rm soft}(q^2_0)$ where $\xi^{\rm Eucl.}_{\rm soft}(q^2_0)$ is given in Eq.~\eqref{eq:softNLL}, and $q^2_0=-3\pm 3$ GeV$^2$. As the parallel form factor must vanish at $q^2=0$ GeV$^2$, we adopt $\xi^{\rm Eucl.}_{\parallel,\,\rm soft}(0)=0$ (i.e.~$q_0^2=0$ GeV$^2$). In addition, as the phase between the resonant and hard-collinear contributions is unknown, we choose to add two arbitrary phases $\delta_\rho$ and $\delta_\omega$ which will be varied between $-\pi/2$ and $\pi/2$.

\subsubsection{Power-suppressed terms}
Having obtained the expression for the leading power and soft or non-perturbative contributions to the form factors,
we now consider the power-suppressed terms. This includes the symmetry breaking corrections and the longitudinal form factor. Considering the leading power suppressed term in the $u$-quark propagator,
we find  
	\begin{align}\label{SB_contri_up}
\Delta F_V^{(u)} =-\Delta F_A^{(u)} =\,&\, Q_u\,f_B\, m_B\, \int_{0}^{\infty} dk_+ \phi_B^+ \frac{1}{q_-^2}
\left(\frac{q^2}{q_-\, k_+ - q^2}+1\right) \nonumber \\
=\,&\, \frac{Q_u}{q_-^2} \left( f_B\, m_B+ q^2\,F_{V/A}^{\rm LO}\right),
   \end{align}
   where $F_{V/A}^{\rm LO}$ can be found in Eq.~\eqref{eq:FVFALO}.
The contribution arising due to the virtual photon emission from the $b$-quark leg is also power suppressed and has been
   neglected while computing the symmetry preserving, leading order contribution to the form factors. Similar to the
   $u$-quark contribution, the emission from the $b$-quark leg yields a symmetry breaking correction which is conveniently
   written as
   \begin{align}\label{SB_contri_bottom}
\Delta F_V^{(b)} =-\Delta F_A^{(b)} =\,&\, \frac{Q_b f_B\,m_B}{q^2-2 m_b\,v.q}.
   \end{align}
  For simplicity, we can therefore define
      \begin{align}\label{asym_contribution}
\Delta F_V^{\rm asym} =F_V^{(u)}+F_V^{(b)}=-(F_A^{(u)}+F_A^{(b)}) .
   \end{align}
    
 Finally the longitudinal form factor can be obtained by matching the second term onto a hadronic matrix element with a transverse derivative acting on the spectator-quark field. This can be computed within the Wandzura-Wilczek approximation, for hard collinear $q^2$:
 \begin{equation}
F_{\parallel}^{\rm NLP}= \int_{0}^{\infty} dk_+ \frac{2 f_B m_B Q_u}{q_-\,k_+-q^2-i\epsilon}\left(\frac{q^2}{m_B^2}\phi_B^-(k_+)-2 \frac{ q_+}{q_-}\phi_B^+(k_+)\right)+\frac{(Q_b-Q_u) f_B}{v\cdot q}.
\end{equation}

\subsubsection{Summary of contributions to the form factors}
Combining Eqs.~\eqref{eq:FVFA-NLL}, \eqref{eq:softLO}, \eqref{eq:softNLL}, \eqref{SB_contri_up} and \eqref{SB_contri_bottom}, we obtain the final 
   results for our form factors  for the decay $B\to \ell\ell\ell'\nu$, 
 \begin{align}
\nonumber  F_V=&F_{V/A}^{\rm NLL}+\xi_V^{\rm NLL}+ \Delta F^{\rm asym}_V\\
\nonumber F_A=&F_{V/A}^{\rm NLL}+\xi_A^{\rm NLL}- \Delta F^{\rm asym}_V\\
\label{eq:FVFA-final} F_{\parallel}=&F_{\parallel}^{\rm NLP}+\xi_\parallel^{\rm NLL}.
 \end{align}
  \begin{figure}[t]
\begin{center}
\includegraphics[width=0.49\textwidth]{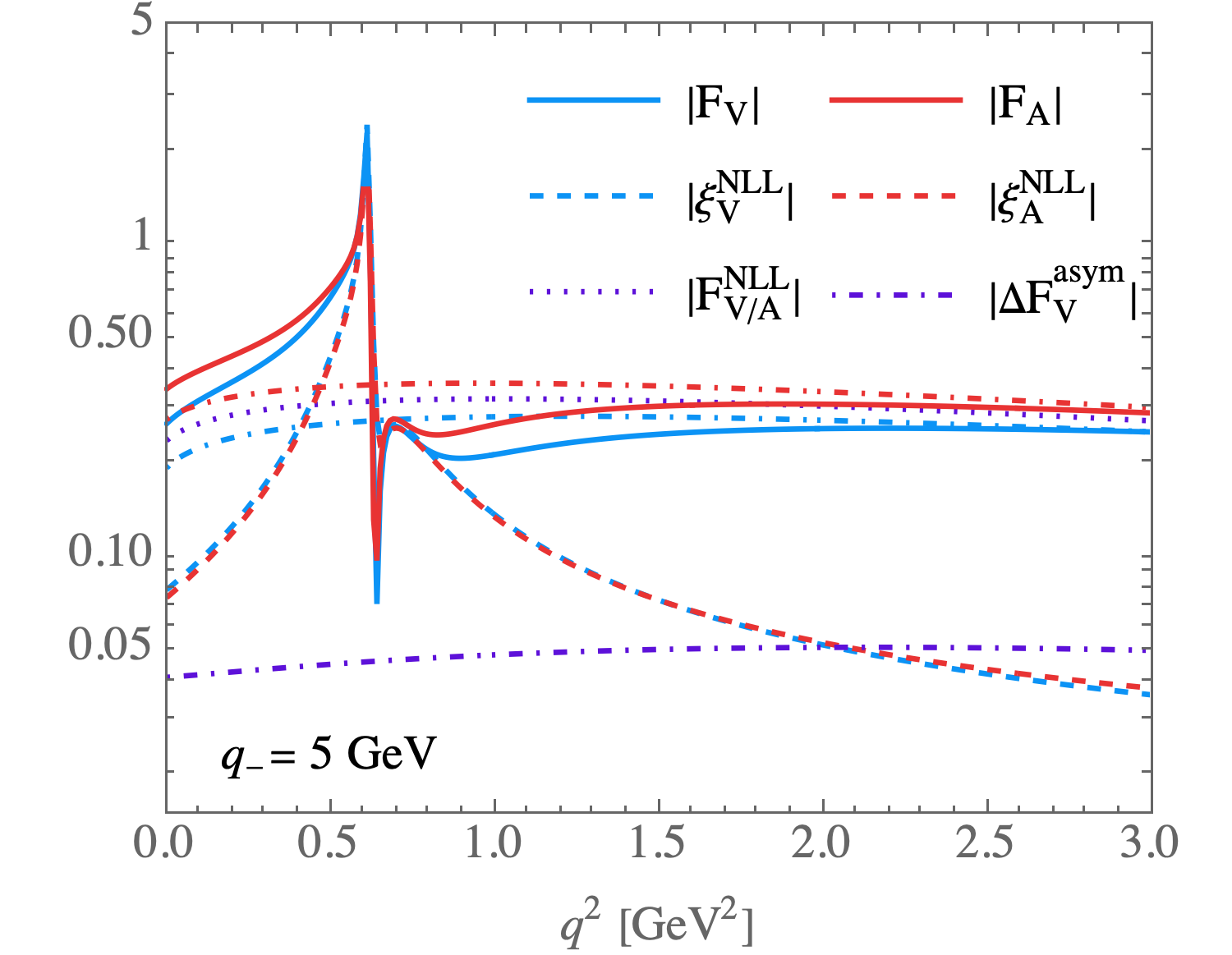} \includegraphics[width=0.50\textwidth]{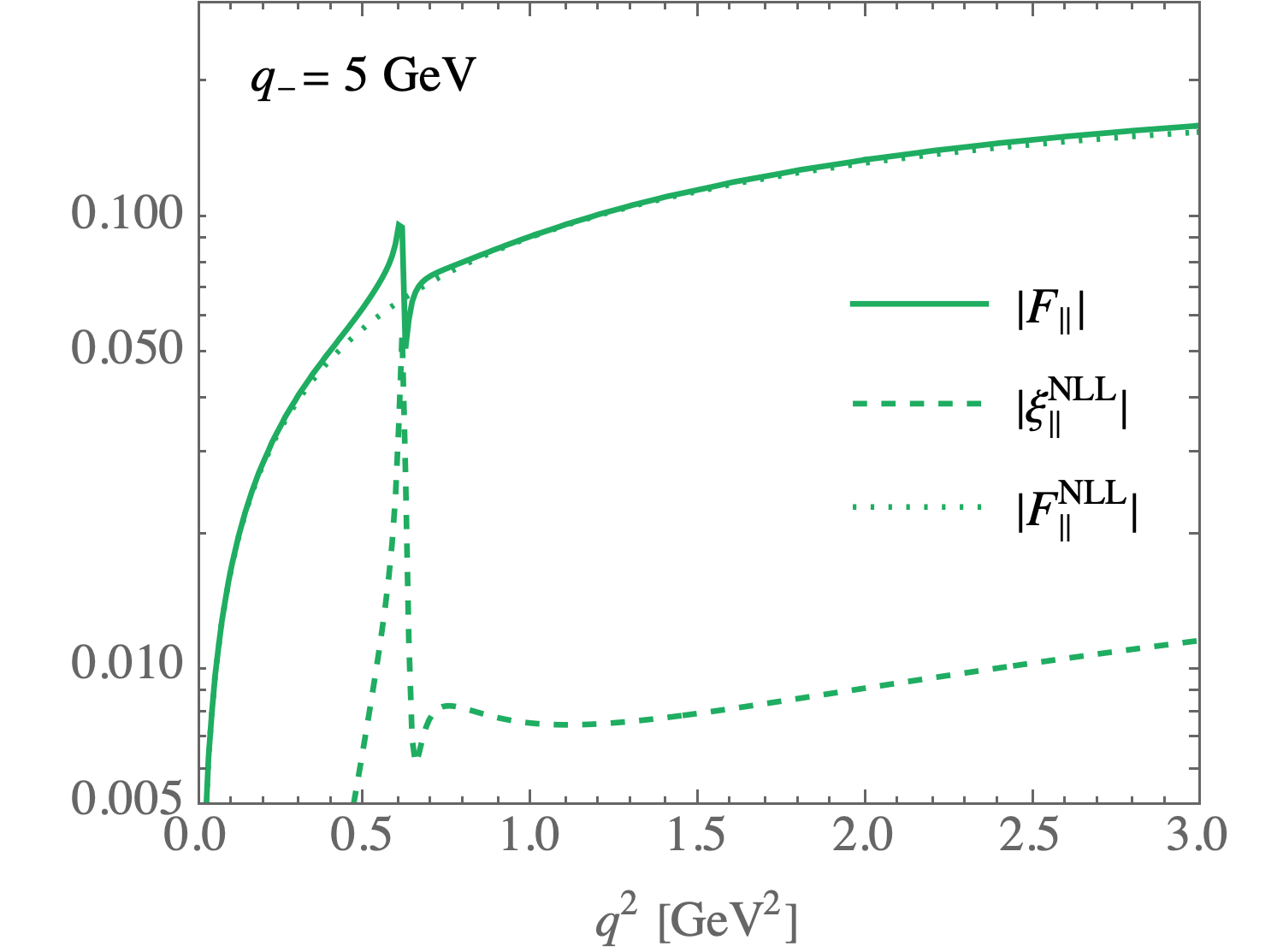}
\caption{\label{fig:formfacbifur} Different contributions to the form factors $F_V$, $F_A$ and $F_\parallel$ are shown as a function of $q^2$ at a fixed value of $q_-=5$ GeV. In the left-hand plot, the blue and red curves show the full form factors $F_V$ and $F_A$ respectively. The red and blue dashed lines represent the soft contributions to the form factors including resonant effects, i.e.~$\xi_{\rm V/A}^{\rm NLL}$. In purple we show the symmetric NLL contribution to the form factors $F_{V/A}^{\rm NLL}$ (dotted) and the antisymmetric contribution $\Delta F^{\rm asym}_{V}$ (dot-dashed). In the right-hand plot, the green lines show $F_\parallel$ (solid) as well as the soft (dashed) and symmetric (dotted) NLL contributions, $\xi_{\rm parallel}^{\rm NLL}$ and $F_\parallel^{\rm NLL}$, as indicated.}
\end{center}
\end{figure}
In order to compare the different contributions to the form factors we plot the individual terms as a function of $q^2$ in Fig.~\ref{fig:formfacbifur} for $q_-=5$ GeV. The blue and red curves show the full form factors at leading order as defined in Eq.~\eqref{eq:FVFA-final}. The dashed blue and red lines show the soft and resonant contributions i.e.~$\xi_{V/A}^{\rm NLL}$ as defined in Eq.~\eqref{eq:softdisp}. 
The purple dotted curves show the symmetric contributions to $F_{V/A}$, $F_{V/A}^{\rm NLL}$ as defined in Eq.~\eqref{eq:FVFA-NLL}. Finally the purple dot-dashed line depicts the sum of the symmetry breaking terms, $\Delta F^{\rm asym}_{V}$  defined in Eq.~\eqref{asym_contribution}. The contribution of photon emission from the $b$ quark is always negative due to the  negative charge of the $b$ quark. We observe that the form factors $F_V$ and $F_A$ are dominated by the symmetric NLL result $F_{V/A}^{\rm NLL}$, except for around the $\rho$ and $\omega$ resonances, where $\xi^{\rm NLL}_{V}$ and $\xi^{\rm NLL}_{A}$ dominate. We find that the resonances contribute over 40\% of the form factors between $q^2=$0.3 and 1.2 GeV$^2$  for $F_V$ and between $q^2=$0.4 and 1.1 GeV$^2$  for $F_A$, meaning that in these regions the sensitivity  to $\lambda_B$ is potentially diminished.  

\begin{figure}[t]
	\begin{center}
		\includegraphics[width=.49\textwidth]{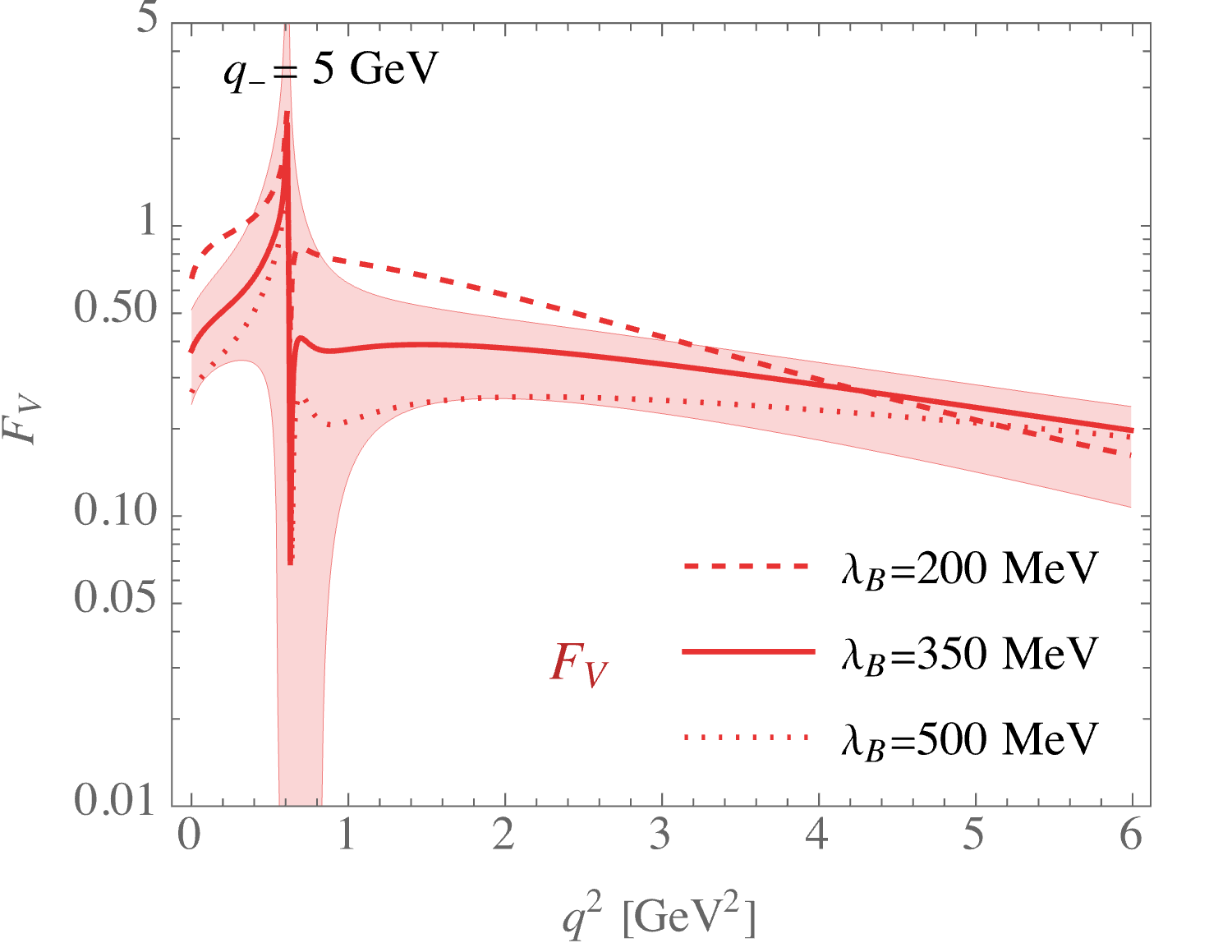}
		\includegraphics[width=.5\textwidth]{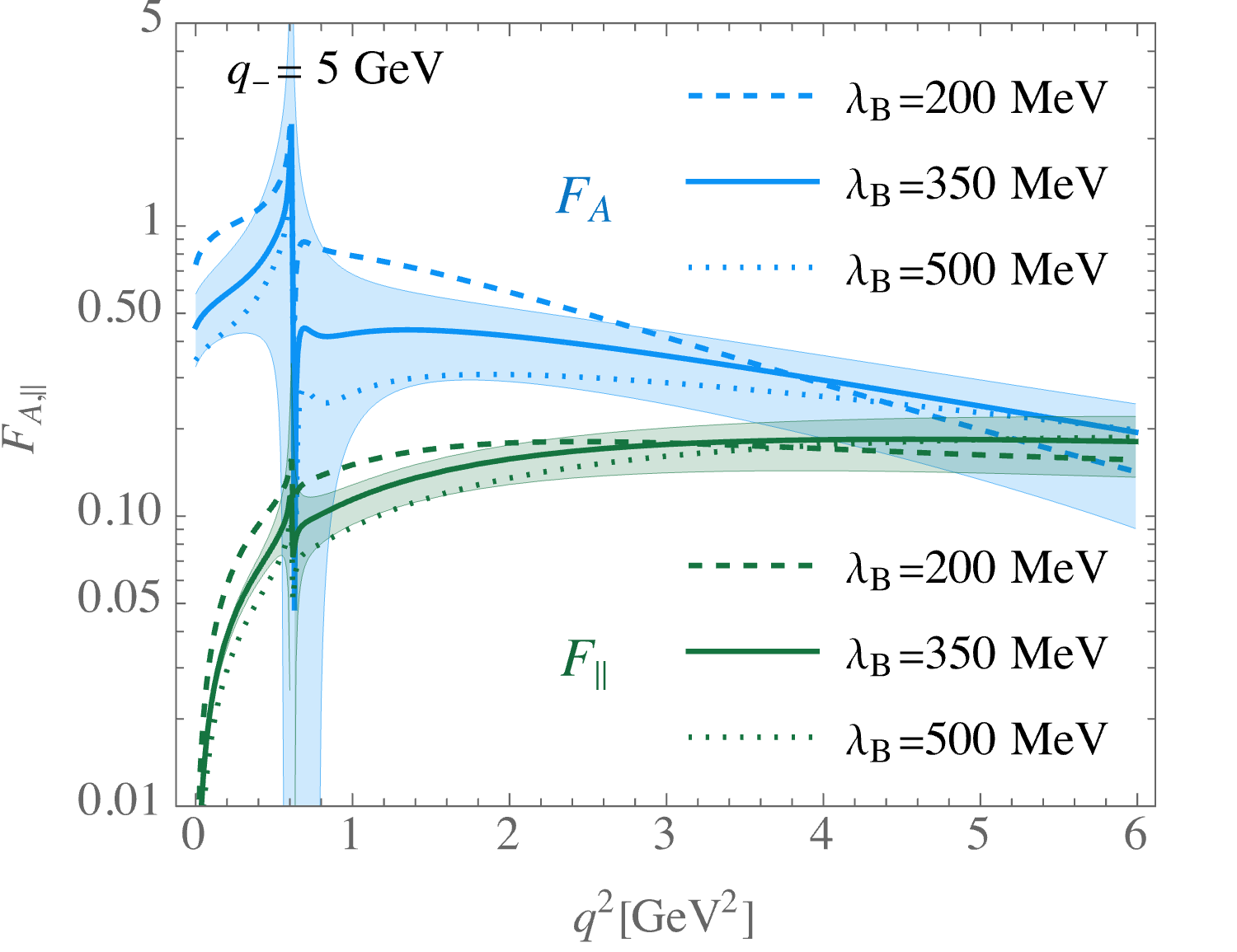}
     \caption{Here we show \label{fig:fvfa} $\left|F_V\right|$ in red (left) and  $\left|F_A\right|$ in blue, $\left|F_\parallel\right|$ in green  (right) as a function of $q^2$ for $q_-=5$ GeV. The solid curves and uncertainty bands correspond to central values of the input parameters and $\lambda_B$ = 350 MeV, the dashed and dotted curves correspond to $\lambda_B$ = 200 and 500 MeV respectively.}
	\end{center}
\end{figure}

In order to compare the uncertainties on the form factors and the dependence on $\lambda_B$, we present the central values (for $\lambda_B(\mu_0)=$350 MeV) of the form factors $F_V$, $F_A$ and $F_\parallel$, as well as the uncertainty bands, shown by the red (left), blue and green (right) bands, in Fig.~\ref{fig:fvfa}, where $\mu_0$ is a reference scale which we set to 1.5 GeV, as discussed in Sec.~\ref{sec:parameters}. The input parameters and the related uncertainties used to calculate this uncertainty band are given in Tab.~\ref{Tableinput} and again will be discussed in detail in Sec.~\ref{sec:parameters}. 
In the same plots, the dashed and dotted lines further show the central values of the form factors $F_V$, $F_A$ and $F_\parallel$ for $\lambda_B=200$ and 500 MeV respectively, and we observe a strong dependence  of the form factors on $\lambda_B$, particularly for lower values of $\lambda_B$. This is in accordance with the results for the branching ratio, which will be discussed in detail in Sec.~\ref{sec:results}.
While the longitudinal form factor is power suppressed, it was pointed out that in the branching ratio it is multiplied with a large factor, such that in the end it contributes similarly to the other form factors~\cite{Beneke:2021rjf}, it is therefore important that it be included.

  Note that the higher twist contributions which have been calculated in Refs.~\cite{Beneke:2018wjp,Wang:2021yrr,Bharucha:2026bwx} were neglected here. 
   These contributions were seen to be relatively small, in Ref.~\cite{Bharucha:2026bwx} the contributions of higher twist corrections to the soft form factors were found to be $\lesssim 5\%$.

\def\arraystretch{1.2}
\begin{table}[t!]
\begin{center}
\begin{tabular}{cccccc}
\toprule
Parameter &Value & Ref.&Parameter &Value & Ref.\\
\midrule
$ m_B$&$5.28~\mathrm{GeV}$ &~\cite{PhysRevD.98.030001} &$ f_B$&$192.0 \pm 4.3$ MeV &~\cite{PhysRevD.98.030001} \\
   $\tau_{B}$&$1.641 \times 10^{-12}$ s &~\cite{PhysRevD.98.030001} &
$\lambda_B$&$ [200-500] $ MeV &~\cite{Beneke:2018wjp}\\ $s_0$&$1.5 \pm 0.1$ GeV$^2$ &~\cite{Beneke:2018wjp}&  $M^2$&$1.25\pm 0.25$ GeV$^2$ &~\cite{Beneke:2018wjp} \\
$m_{\rho}$&$0.775$ GeV &~\cite{PhysRevD.98.030001}  
 & $m_{\omega}$&$0.782$ GeV &~\cite{PhysRevD.98.030001}  \\
 $\Gamma_{\rho}$&$147.8$ MeV &~\cite{PhysRevD.98.030001}  
 & $\Gamma_{\omega}$&$8.49$ MeV &~\cite{PhysRevD.98.030001}  \\
 $f_{\rho}$&$213$ MeV &~\cite{Bharucha:2015bzk}  
 & $f_{\omega}$&$197$ MeV &~\cite{Bharucha:2015bzk}  \\
$\Lambda_{\rm QCD}^{n_f=3}$&$0.33965$ GeV &\cite{PhysRevD.98.030001,Chetyrkin:2000yt}&  $\left|V_{ub}\right|^{\text{excl}}$&$(3.70\pm 0.16)\times  10^{-3}$&~\cite{PhysRevD.98.030001} \\
 $m_{\mu}$&$0.105 ~\mathrm{GeV}$ &~\cite{PhysRevD.98.030001}& $m_e$&$ 0.511 \times 10^{-3}~\mathrm{GeV}$ &~\cite{PhysRevD.98.030001} \\
$\alpha_{\rm em}(m_b)$&$1/132$ &~\cite{Pivovarov:2000cr} & $G_F$&$1.166 \times 10^{-5}
 ~\mathrm{GeV}^{-2}$&~\cite{PhysRevD.98.030001}  \\
\bottomrule
\end{tabular}
\caption{Numerical values of parameters adopted in our analysis along with the associated uncertainties (where relevant) and the references from which they were taken. \label{Tableinput}}
\end{center}
\end{table}

	\section{Numerical analysis and results}
	\label{sec:results}
In this section, we will first outline the kinematics required to describe the differential decay distribution of the four leptonic decay of charged $B$ meson. This will be followed by a discussion of the numerical parameters implemented in our analysis. We will then present the results, where numerical predictions for the decay process under study are provided in specific $q^2$ bins, along with a thorough analysis of the uncertainties.
	
	\subsection{Kinematics}
	We follow the definitions in \cite{Pais:1968zza} for the kinematics. In order to examine the  partial decay rate of $B^+ (p_B) \to \ell^+(q_1) \ell^-(q_2)\ell^{\prime+}(p_1)\nu(p_2)$, it is useful to introduce the following combinations  of the final state particles' four-momenta:
	\begin{align}
	q=q_1+q_2; & & Q=q_1-q_2& & p=p_1+p_2 & & P=p_1-p_2. 
	\end{align}
	This four-body partial decay rate can then be described via five independent variables:
	\begin{itemize}
	\item  the effective mass squared, $p^2$ and $q^2$, of  the $\ell'\nu$ and $\ell^+\ell^-$ system respectively,
	\item the angles $\theta_{\gamma}$ of the $\ell^+$ in the $\ell^+\ell^-$ center-of-mass system with respect to the $\ell^+\ell^-$ line of flight in $B$ rest frame, and $\theta_W$ of the $\ell'$ in the $\ell'\nu$ center-of-mass system with respect to the $\ell'\nu$ line of flight in $B$ rest frame, where
	\begin{equation}
	\cos\,\theta_{\gamma} = -\frac{\vec{Q}~.~\vec{p}}{|\vec{Q}|~|\vec{p}|},\qquad\mbox{and}\qquad 
	\cos\,\theta_W = -\frac{\vec{q}~.~\vec{P}}{|\vec{q}|~|\vec{P}|},
	\end{equation}
	\item the angle $\phi$ between the planes of the two lepton pairs,
	\begin{equation}
	\sin\phi = \frac{(\vec{q}\times \vec{P}) \times( \vec{p}\times \vec{Q})}{|\vec{q}\times \vec{P}|~| \vec{p}\times \vec{Q}|}.
	\end{equation}
		\end{itemize}
\noindent In terms of these five variables, the partial decay distribution is given by,
\begin{equation}\label{partial}
d^5\Gamma= \frac{\pi \lambda(m_B^2,q^2,p^2)^{1/2}}{( 4\pi)^7}m_B^5|A|^2  d\Phi
\end{equation}
where $d\Phi$ denotes an element of the four-body phase space, and is defined via
\begin{equation}
d\Phi =dq^2~dp^2~d\cos\theta_{\gamma}~d\cos\theta_W~d\phi,
\end{equation}
$A$ is the amplitude of the process defined in Eq.~\eqref{eq:ampli} and the K\"{a}ll\'{e}n function is given by
\begin{equation}
\lambda(a,b,c)=a^4+b^4+c^4-2(a^2b^2+b^2c^2+c^2a^2).
\end{equation}
Note that the decay distribution in Eq.~\eqref{partial} corresponds to the case when $\ell\neq\ell^{\prime}$. If the leptons in the final state  are all of the same flavour, there is an additional contribution to the  amplitude due to the possible exchange of same-sign leptons.  The expression for the partial decay rate given in Eq.~\eqref{partial}, is then modified as $A\to \mathcal{M}\equiv A-A'$, resulting in,
\begin{equation}
\left|\mathcal{M}\right|^2=  \frac{1}{2}\Big[\left|A\right|^2 d\Phi +\left|A'\right|^2 d\Phi' - (A A'^{\dagger}+A^{\dagger}A')d\Phi\Big],
\end{equation}
where $A^{\prime}$ and $\Phi^{\prime}$ are obtained from $A$ and $\Phi$ respectively $p_1\leftrightarrow k_2$. 

To obtain branching ratios, we first  integrate over all variables except $q^2$ in the ranges 
		\begin{eqnarray}
	\nonumber	m_{\ell'}^2 \leq p^2\leq (m_B-\sqrt{q^2})^2\\
		0\leq \theta_\gamma,\theta_W\leq \pi, ~ 0\leq \phi\leq 2 \pi.
		\end{eqnarray}
While the physical range of $q^2$ is given by $4 m_{\ell}^2 \leq q^2\leq (m_B- m_{\ell'})^2$, since the factorization is only valid for low $q^2$, we will present binned branching ratios in the $q^2$ bins, $[4m_{\mu}^2,0.96\,\,\mathrm{GeV}^2]$ and $[1.1,3]$ GeV$^2$.

\subsection{Parameters}
\label{sec:parameters}

Before coming to the results, a discussion of our choices for the numerical input parameters is in order. The critical hadronic input parameters for our analysis include the leading and sub-leading 2-particle light-cone distribution amplitude $\phi_B^+$ and $\phi_B^-$, the continuum threshold $s_0$,  Borel Parameter $M^2$, and the decay constant of $B$ meson $f_B$. In addition, the hard-collinear scale $\mu$ and the CKM matrix element $V_{ub}$ play an important role. 
Starting with $\phi_B^+$ and $\phi_B^-$, there exist several possible choices for the parametrisation in the literature (see for example: \cite{Braun:2012kp,DescotesGenon:2001hm,DescotesGenon:2002mw}).
We choose the form of $\phi_B^+$ at a reference scale $\mu_0=1.5$ GeV to be (see e.g.~Ref.~\cite{Braun:2012kp}),
\begin{equation}
\phi_B^+(k,\mu_0)=\frac{k}{\lambda_B(\mu_0)^2}e^{-k/\lambda_B(\mu_0)}
\end{equation}
where, $\lambda_B(\mu_0)$ is the first inverse moment of the $B$-meson DA defined in Eq.~\eqref{eq:lambdaB}, at the scale $\mu_0$.
The scale dependent $\phi_B^+(k,\mu)$ is then obtained following Appendix A of Ref.~\cite{Beneke:2018wjp}.
In order to obtain the subleading 2-particle DA $\phi_B^-$ from $\phi_B^+$, we make use of the Wandzura-Wilczek approximation~\cite{Beneke:2000wa}
\begin{equation}
\phi_B^-(k,\mu)=\int_k^\infty \frac{dk'}{k'}\phi_B^+(k',\mu).
\end{equation}
Our results clearly depend on the choice of $\lambda_B(\mu_0)$, for which we adopt the range given in Tab.~\ref{Tableinput}, following Ref.~\cite{Beneke:2018wjp}.

For the continuum threshold and the Borel parameter we take the standard values as advocated in Refs.~\cite{Braun:2012kp, Beneke:2018wjp}.
The final hadronic parameter is the $B$-meson decay constant $f_B$, for which we adopt the average found in Ref.~\cite{PhysRevD.98.030001}.
$V_{ub}$ is the dominant source of uncertainty after $\lambda_B$. As this is an exclusive decay, we feel that it is appropriate to make use of the exclusive average as an input parameter, as found in Ref.~\cite{PhysRevD.98.030001}.
While the $b$-quark mass does not have a large impact on our results, we choose to use the pole mass as input, adopting conservative errors $4.8\pm0.1$ GeV.
Finally the form factors $V(p^2)$ and $A_1(p^2)$ for the $B\to\rho$ and $B\to\omega$ transitions along with the uncertainties on these are taken from Ref.~\cite{Bharucha:2015bzk}.
A summary of the values of the parameters used in our analysis can be found in Tab.~\ref{Tableinput}.


\subsection{Results}
 We now present numerical predictions for the form factors and branching fractions for $B\to \ell\ell\ell^{\prime}\nu$ in the $q^2$ (GeV$^2$) bins: $[4m_{\ell}^2,0.96$ GeV$^2$] and $[1.1,3]$ GeV$^2$ within the updated theoretical framework described in Sec.~\ref{sec:SymNLL}. Unless otherwise stated, all predictions include the complete set of form factors, the NLL leading-power corrections, the $\mathcal{O}(\alpha_s^0)$ next-to-leading-power contributions, and the soft corrections obtained from the once-subtracted dispersion relations.
 In order to understand the origin of the uncertainties on the result, we study the different contributions to the uncertainty for $B\to \mu\mu e\nu$ (for $\lambda_B=350$ MeV):
\begin{align}\label{Result:numerics5}
\nonumber \hspace{-.1cm}10^8\,\mathcal{B}\big|_{ [4\,m_\mu^2,\,0.96\, \mathrm{GeV}^2]} = 
\,1.56\,\Bigg[1+&\begin{footnotesize}
\begin{pmatrix}
+0.034\\
-0.057
\end{pmatrix}_{\mu}+
\begin{pmatrix}
+0.088\\
-0.085
\end{pmatrix}_{|V_{ub}|}+
\begin{pmatrix}
+0.030\\
-0.029
\end{pmatrix}_{f_B}
+\end{footnotesize}\\
& \begin{footnotesize}
\begin{pmatrix}
+0.340\\
-0.182
\end{pmatrix}_{\delta_\rho}+\begin{pmatrix}
+0.136\\
-0.132
\end{pmatrix}_{\delta_\omega}+
\begin{pmatrix}
+0.096\\
-0.009
\end{pmatrix}_{q_0^2}\end{footnotesize}\Bigg]\,,\\
\nonumber \hspace{-.1cm}10^9\,\mathcal{B}\big|_{ [\,1.1,\,3.0]\, \mathrm{GeV}^2} = 
\,2.37\,\Bigg[1+&\begin{footnotesize}
\begin{pmatrix}
+0.437\\
-0.659
\end{pmatrix}_{\mu}+
\begin{pmatrix}
+0.884\\
-0.846
\end{pmatrix}_{|V_{ub}|}+
\begin{pmatrix}
+0.427\\
-0.418
\end{pmatrix}_{f_B}
+\end{footnotesize}\\
&\begin{footnotesize}
\begin{pmatrix}
+2.264\\
-0.886
\end{pmatrix}_{\delta_\rho}+ \begin{pmatrix}
+0.672\\
-0.294
\end{pmatrix}_{\delta_\omega}+
\begin{pmatrix}
+1.188\\
-0.084
\end{pmatrix}_{q_0^2}\end{footnotesize}\Bigg]\,,
\end{align}
\noindent
and also for the case of $B\to ee\mu\nu$:
\begin{align}\label{Result:numerics6}
\nonumber \hspace{-.1cm}10^8\,\mathcal{B}\big|_{ [0.01,\,0.96]\, \mathrm{GeV}^2} = 
\,2.09\,\Bigg[1+&\begin{footnotesize}
\begin{pmatrix}
+0.033\\
-0.052
\end{pmatrix}_{\mu}+
\begin{pmatrix}
+0.088\\
-0.085
\end{pmatrix}_{|V_{ub}|}+
\begin{pmatrix}
+0.032\\
-0.031
\end{pmatrix}_{f_B}
+\end{footnotesize}\\
& \begin{footnotesize}
\begin{pmatrix}
+0.313\\
-0.182
\end{pmatrix}_{\delta_\rho}+\begin{pmatrix}
+0.118\\
-0.119
\end{pmatrix}_{\delta_\omega}+
\begin{pmatrix}
+0.129\\
-0.012
\end{pmatrix}_{q_0^2}\end{footnotesize}\Bigg]\,,\\
\nonumber \hspace{-.1cm}10^9\,\mathcal{B}\big|_{ [\,1.1,\,3.0]\, \mathrm{GeV}^2} = 
\,2.40\,\Bigg[1+&\begin{footnotesize}
\begin{pmatrix}
+0.437\\
-0.659
\end{pmatrix}_{\mu}+
\begin{pmatrix}
+0.884\\
-0.846
\end{pmatrix}_{|V_{ub}|}+
\begin{pmatrix}
+0.427\\
-0.418
\end{pmatrix}_{f_B}
+\end{footnotesize}\\
&\begin{footnotesize}
\begin{pmatrix}
+2.265\\
-0.888
\end{pmatrix}_{\delta_\rho}+ \begin{pmatrix}
+0.672\\
-0.295
\end{pmatrix}_{\delta_\omega}+
\begin{pmatrix}
+1.186\\
-0.084
\end{pmatrix}_{q_0^2}\end{footnotesize}\Bigg]\,.
\end{align}
We find that the largest contribution to the uncertainty, of up to $15\%$ arises from the phases of the resonant contributions, $\delta_\rho$ and $\delta_\omega$. Another major source of uncertainty is the CKM matrix element $|V_{ub}|$, contributing at the $8\%$ level. 
  Since $|V_{ub}|$ enters as an overall factor in the branching ratio, the resulting uncertainty is independent of the phase space. The hard-collinear factorization scale $\mu$ also provides a non-negligible source of uncertainty. The uncertainty associated with the subtraction point provides a direct estimate of the residual model dependence of the dispersive treatment. As illustrated in Eqs.~\eqref{Result:numerics5} and \eqref{Result:numerics6}, this uncertainty remains subdominant compared with those associated with the resonance phases and $|V_{ub}|$.
  \def\arraystretch{1.2}
\begin{table}[t!]
\begin{center}
\begin{tabular}{ccc}
\toprule
 $q^2$ bin (GeV$^2$) & $10^8\,\mathcal{B}(B\to \mu\mu e\nu)$ &  $10^8\, \mathcal{B}(B\to ee\mu\nu)$\\
\midrule

 $[q_1^2,0.96]$ & $2.09\left(^{+0.37}_{-0.24}\right)$ GeV$^2$ &$1.56\left(^{+0.39}_{-0.25}\right)$ GeV$^2$\\
 $[1.1,3.0]$ & $0.24\left(^{+0.29}_{-0.15}\right)$ GeV$^2$ &$0.24\left(^{+0.29}_{-0.15}\right)$ GeV$^2$\\
 \bottomrule
\end{tabular}
\caption{\label{tab:resultsBlllnu} Binned branching ratios for $B\to \mu\mu e\nu$ and $B\to ee\mu \nu$ in two $q^2$ bins as indicated, where $q_1^2=4 m_\mu^2$ for $B\to \mu\mu e\nu$ and 0.01 GeV$^2$ for $B\to e e\mu \nu$. We do not include the region  $[0.96,1.1]$ GeV$^2$ in order to decrease the sensitivity to the $\phi$ pole.}
\end{center}
\end{table}

Note that an interesting consequence of lepton flavour universality (LFU) in the SM is the branching ratios for different lepton pairs, measured in the same kinematic range, should turn out to be equal, 
\begin{align}\label{Eq:LFU}
\mathcal{B}(B\to \mu\mu e\nu)\big|_{[q^2_{\rm low},q^2_{\rm high}]}& =\mathcal{B}(B\to ee \mu \nu)\big|_{[q^2_{\rm low},q^2_{\rm high}]}.
\end{align}
In this study, we found that the relation holds up to the accuracy of our results, as seen in Tab.~\ref{tab:resultsBlllnu}. As the lower bin ranges differ for the two decays, this can more easily be seen for the higher $q^2$ bin in Tab.~\ref{tab:resultsBlllnu}.

\begin{figure}
\begin{center}
\hspace{-.5cm}
\includegraphics[width=.5\textwidth]{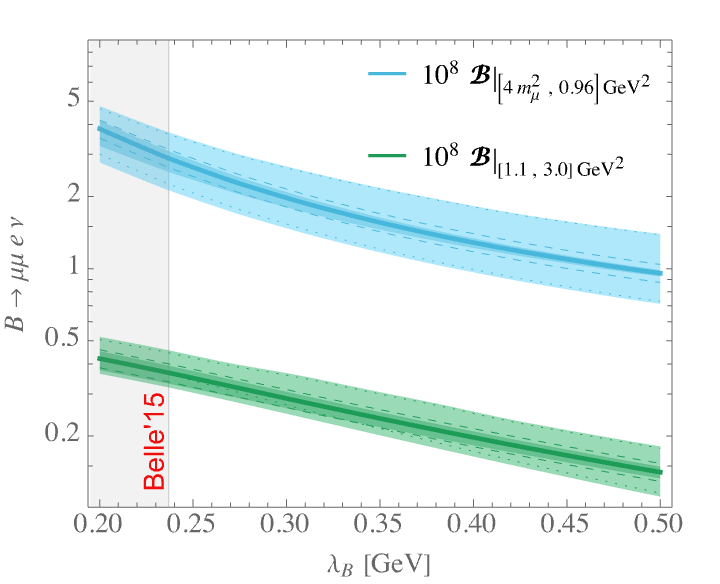}\hspace{-.1cm} \includegraphics[width=.5\textwidth]{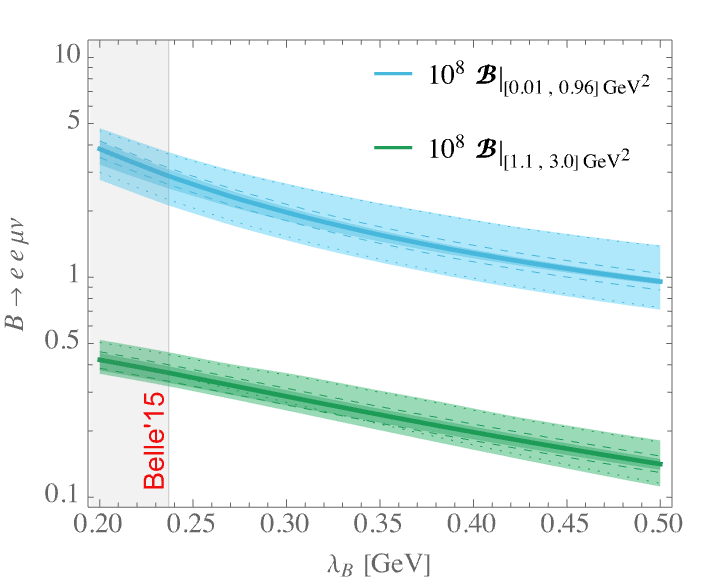} 
\caption{ \label{fig:brvslamB}The integrated branching ratio of $B\to \mu\mu e \nu$ and $B\to ee\mu \nu$ in the low $q^2$ bins $[4\, m_\mu^2,0.96$ GeV$^2$] and $[0.01,0.96$ GeV$^2$] respectively and the high  $q^2$ bin $[1.1,3.0]$ GeV$^2$ is shown as a function of $\lambda_B$. The solid line shows the central value, the light-shaded band indicates the total uncertainty, and the dark-shaded band indicates the uncertainty due to all parameters except the subtraction point $q^2_0$ and the phases of the resonances. Note that uncertainties are added in quadrature. The dashed and dotted lines show the uncertainty coming from $q^2_0$ and the phases of the resonant contributions respectively. The vertical grey band corresponds to the exclusion on $\lambda_B$ at 90\% C.L.~from Belle~\cite{Heller:2015vvm}. }
\end{center}
\end{figure}

In the above discussion of the uncertainties, we have not yet mentioned the input parameter $\lambda_B$. The reason is that this is the very parameter we propose to measure or constrain via measurements of the branching ratios. 
Such a measurement relies critically on the dependence of the integrated branching ratio on $\lambda_B$, which we show in Fig.~\ref{fig:brvslamB}. Here on the left one sees the branching ratio for the decay $B\to \mu\mu e\nu$ in the bin $[4m_{\mu^2},0.96$ GeV$^2$] (blue) and $[1.1,3.0$ GeV$^2$] (green) as a function of $\lambda_B$, where the central value is shown by the solid line, the total uncertainty is indicated by the light shaded band and the uncertainty band due to all parameters except the subtraction point $q^2_0$,  and the resonance phases $\delta_\rho$ and $\delta_\omega$ is shown by the dark shaded band. This demonstrates the large contribution to the uncertainty of the phases of the resonances. The dashed and dotted lines show the uncertainty associated with $q^2_0$ and the phases of the resonant contributions respectively. We further show the lower bound $\lambda_B>238$ MeV obtained by Belle mentioned in Sec.~\ref{sec:results}~\cite{Heller:2015vvm}. On the right we plot the analogous plot for the decay $B\to ee\mu\nu$, the only difference being that a cut-off is imposed for the lower bin in the branching ratio of 0.01 GeV$^2$. We see that the uncertainty from the phases $\delta_\rho$ and $\delta_\omega$ dominate, particularly at larger values of $\lambda_B$. For both decays, there is clearly a marked dependence of the binned branching ratio on $\lambda_B$, more so for the lower  $q^2$ bin.  We note that for the higher $q^2$ bin, while the dependence on $\lambda_B$ is less marked, the smaller effects of the resonances mean that the uncertainties are reduced.
Ultimately, which bin provides the greater sensitivity will depend on the experimental precision. The final question to be answered is therefore whether the experiments LHCb and Belle II can measure these binned branching ratios, and if so with what accuracy.
 
  Compared to the original version of this work, the most significant numerical changes arise from the improved treatment of the soft contributions and the inclusion of the longitudinal form factor. While these modify the individual form factors, they leave the overall dependence of the branching fractions on $\lambda_B$ largely unchanged.
Our numerical predictions are compatible with the existing QCD-factorization analyses in the low-$q^2$ region, an extensive numerical comparison has been performed with Ref.~\cite{Beneke:2021rjf}. The present work complements these studies by providing phenomenological predictions based on the updated LCSR and dispersive treatment of the soft contributions.

Here we stress again that experimentally, $q^2$ can only be defined uniquely for the case $\ell\neq\ell'$. Therefore the measurement of the integrated branching ratio in the $q^2$ bin we advocate can more easily be defined for this case.\footnote{In Ref.~\cite{Beneke:2021rjf}, it was shown that by making certain kinematic cuts it should be possible to theoretically interpret measurements of the case $\ell=\ell'$.}  As mentioned in Sec.~\ref{sec:intro}, so far the only limit on $B\to \ell\ell\ell'\nu$ decays available is from the LHCb experiment for the case $\ell=\ell'=\mu$, where with 4.7 fb$^{-1}$ they obtain an upper limit on the branching ratio of $1.6\times 10^{-8}$ at 95\% C.L., in the region where the lowest of the two $\mu^+\mu^-$ mass combinations is below 0.98 GeV. As our formalism is only valid for low $q^2$, we cannot compare our results to this prediction.  The sensitivity of LHCb for the case $\ell\neq\ell'$ has not yet been made public, but given the existing limit for $\ell=\ell^\prime=\mu$, and taking a conservative guess that the yield would diminish by a factor 3-4, the prospects for this channel with the 50 ab$^{-1}$ expected with the full Upgrade I data set, let alone the 300 ab$^{-1}$ at the end of Upgrade II, look very promising. For Belle II, the measurement of $B\to\gamma\ell\nu$ would probably provide a more precise measurement of $\lambda_B$, given the branching ratio is $\mathcal{O}(10)$ times larger~\cite{Kou:2018nap}. 
With the full Belle II data set of 50 ab$^{-1}$, a factor $\mathcal{O}(10)$ reduction in the statistical uncertainty should be possible, more details can be found in Ref.~\cite{Kou:2018nap}. However, a measurement of the partial  branching fraction for $B\to ee\mu\nu$ and $B\to \mu\mu e\nu$, in the low $q^2$ bin $[0.01,0.96]$ GeV$^2$ and $[4 m_\mu^2,0.96$ GeV$^2$] respectively, could provide additional interesting information.

\section{Conclusions }
\label{sec:conclusion}
In this paper we have studied the purely leptonic decay modes $B\to \ell\ell\ell'\nu$ for $\ell,\ell'=e,\mu$ in the low $q^2$ region at NLL, including the leading  $1/m_b$ and $1/q^2$ corrections, as well as the soft corrections at NLL in LCSR.
This work is motivated by the possibility to measure $\lambda_B$ at LHCb, since a systematic theoretical study with uncertainties for these four-lepton modes was lacking, and in light of the recent results from LHCb~\cite{Aaij:2018pka}.
The present analysis provides an updated phenomenological study within the current state-of-the-art theoretical framework.
We have provided a numerical comparison of the various contributions to the form factors $F_V$, $F_A$ and $F_\parallel$ in Fig.~\ref{fig:formfacbifur}, where we see the impact of the resonances on the form factors at $q^2\sim$0.6 GeV$^2$. 
Note that certain higher twist contributions which were calculated in the state-of-the-art $B\to\gamma\ell\nu$ analysis have been neglected here, i.e.~the $1/m_b$ and $1/E_\gamma$ higher twist corrections and the twist 3 to 6 contributions to the soft correction. 
These were calculated in Refs.~\cite{Beneke:2021rjf,Wang:2021yrr,Bharucha:2026bwx}, and their effect is of the order $\lesssim 5\%$, negligible compared to the uncertainty coming from $|V_{ub}|$. 

Relative to the original version of this work, the theoretical framework has been substantially updated to incorporate developments in the literature.
In particular, we now determine the soft contributions using once-subtracted dispersion relations where the subtraction point in the Euclidean corresponds to the LCSR results~\cite{Braun:2012kp}.
These improvements place the calculation on the same theoretical footing as the current state-of-the-art treatment of $B\to\gamma^*$ form factors while preserving the main phenomenological conclusions regarding the sensitivity of $B\to\ell\ell\ell'\nu$ to $\lambda_B$.
As in Refs.~\cite{Beneke:2021rjf} and \cite{Wang:2021yrr}, our form factors are not expressed in the basis proposed in Refs.~\cite{Kurten:2022zuy,Bharucha:2026bwx}, this is worthwhile pursuing in the future.

We advocate the measurement of the partial branching fractions for $B\to \mu\mu e\nu$ and $B\to ee\mu\nu$ in the low $q^2$ bins $[4 m_\mu^2,0.96$ GeV$^2$] and $[0.01,0.96$ GeV$^2$] respectively as well as the intermediate bin $[1.1,3]$ GeV$^2$. 
We find that the uncertainty is at the 40\% level for the low $q^2$ bin and 30\% for the higher $q^2$ bin, dominant contributions coming from the unknown phases of the resonant contributions with respect to the non-resonant amplitude and from $|V_{ub}|$.
We further show the dependence of the partial branching ratio on $\lambda_B$ in Fig.~\ref{fig:brvslamB}, and find that the dependence far outweighs the remaining uncertainties, suggesting that given a value of the partial branching ratio a measurement of $\lambda_B$ should be feasible.
While the lower $q^2$ bin is more sensitive to $\lambda_B$, it also suffers larger uncertainties due to the resonances, such that the experimental sensitivities in these two bins will ultimately determine which provides the stronger constraint on the structure of the $B$ meson.
While there are no official projections for these channels at LHCb and Belle II, simple estimates show that the prospects to measure the partial branching ratio is promising. 
We therefore look forward to these results and to the potential measurement of $\lambda_B$, complementary to that of $\mathcal{B}(B\to\gamma\ell\nu)$ at Belle II.

	\section*{Acknowledgements}
We thank Martin Beneke, J\'{e}r\^{o}me Charles and Yao Ji for useful discussions, as well as Racha Cheaib, Francesco Polci, Justine Serrano, and William Sutcliffe for important input concerning the potential experimental sensitivities. AB and BK are further grateful for their time spent at the Institute of Nuclear Theory, Seattle, attending the Heavy-Quark Physics and Fundamental Symmetries program (INT-19-2b), during which important progress on the project was made. The work by NM at the Physical Research Laboratory is supported by the Department of Space (DoS), Government of India. NM also acknowledges
the partial support under the MATRICS project (MTR/023/000442) from the Science $\&$ Engineering
Research Board (SERB), Department of Science and Technology (DST), Government of India.

\appendix
\section{Hadronic matrix element}\label{AppendixA}
Let us define the hadronic matrix element, entering in the amplitude given in Eq.~\eqref{eq:ampli}, by
	\begin{equation}\label{HME}
	T_{\mu\rho}(p,q)=i \int d^4x e^{iqx} \left<0|T\{j_{\mu}^{em}(x)\bar{u}\Gamma_{\rho}b(0)\}|B(p+q)\right>.
	\end{equation}
The most general decomposition of this hadronic matrix element is given by, 
\begin{equation}\label{generaldecomposition}
T_{\mu\rho}=a\,g_{\mu\rho}+b\,q_\mu p_\rho + c\,p_{\mu}q_{\rho} + d \,\epsilon_{\rho\mu\lambda\sigma}p^{\lambda}q^{\sigma}+e\,p_\mu p_\rho+f\,q_\mu q_\rho.
\end{equation}
 	Constraints on $T_{\mu\rho}$ can be obtained using the Ward identity, i.e.~ the current conservation of the electromagnetic current, $\partial^{\mu}j_{\mu}^{em}=0$. This is implemented by differentiating the correlation function in the definition of $T_{\mu\rho}$, which gives, 
	\begin{equation}\label{WardID}
	q^{\mu}T_{\mu\rho} =i (p+q)_{\rho}f_B.
	\end{equation}
Applying Eq.~\eqref{WardID} to Eq.~\eqref{generaldecomposition}, this provides two constraints on the coefficients:
\begin{align}
a+c\,p\cdot q+f\,q^2=& if_B\\
b\,q^2+e\,p\cdot q=& if_B.
\end{align}
In addition, terms containing $p_{\rho}$ and $q_\mu$ can be neglected since these contributions will always be proportional to the mass of lepton. 
We therefore arrive at the simplified expression
\begin{equation}
T_{\mu\rho}=    a\,g_{\mu\rho} +\frac{if_B-a-f\,q^2}{p\cdot q}q_\rho p_\mu +d\,\epsilon_{\rho\mu\lambda\sigma}p^\lambda q^\sigma.
\end{equation}
From this simplified expression, on replacing $p$ by $m_B v-q$, we can identify three independent form factors,
 \begin{equation}
T_{\mu\rho}=i F_A(g_{\mu\rho}\,v\cdot q-v_\mu q_\rho)+F_V\epsilon_{\rho\mu\lambda\sigma} v^\lambda q^\sigma-i F_{\parallel}v_\mu q_\nu.
\label{finaldef}
\end{equation}

\printbibliography
\end{document}